\documentclass[hidelinks,prd,eqsecnum,amsmath,amssymb,nofootinbib,12pt,superscriptaddress]{revtex4-1}

\usepackage{dcolumn}
\usepackage{bm}
\usepackage{xcolor}
\usepackage[applemac]{inputenc}
\usepackage[spanish,english]{babel}
\usepackage{amsmath,amssymb,amsfonts,latexsym,cancel,amsthm,hyperref,bbm}
\usepackage{graphicx}
\usepackage{color}
\usepackage{soul}
\usepackage{ulem}
\usepackage{slashed}
\usepackage{braket}
\usepackage[mathscr]{euscript}
\usepackage{physics}
\usepackage{mathrsfs}
\usepackage{pgfplots}
\usepackage{float}
\usepackage{xfrac}
\usepackage{mathtools}

\allowdisplaybreaks

\newcommand{\la}{\langle}
\newcommand{\ra}{\rangle}

\newcommand{\be}{\begin{equation}}
\newcommand{\ee}{\end{equation}}
\newcommand{\bea}{\begin{eqnarray}}
\newcommand{\eea}{\end{eqnarray}}
\newcommand{\bes}{\begin{subequations}}
\newcommand{\ees}{\end{subequations}}

\begin{document}

    \title{\bf Linear Response Analysis of the Semiclassical Approximation to Spin $\sfrac{1}{2}$ Quantum Electrodynamics in 1+1 Dimensions}
    
    \author{Ian M. Newsome}\email{newsim18@wfu.edu}
    \affiliation{Department of Physics, Wake Forest University, Winston-Salem, NC, 27109, USA}

    \author{Paul R. Anderson}\email{anderson@wfu.edu}
    \affiliation{Department of Physics, Wake Forest University, Winston-Salem, NC, 27109, USA}

    \author{Eric M. Grotzke}\email{eric@minimal.email}
    \affiliation{Department of Physics, Wake Forest University, Winston-Salem, NC, 27109, USA}

    \begin{abstract}
        An investigation of the validity of the semiclassical approximation to quantum electrodynamics in 1+1 dimensions is given. The criterion for validity used here involves the impact of quantum fluctuations introduced through a two-point function which emerges naturally when considering the stability of the backreaction equation to linear order perturbations, resulting in the linear response equation. Consideration is given to the case of a spatially homogeneous electric field generated by a classical source, coupled to a quantized massive spin $\sfrac{1}{2}$ field. Solutions to the linear response equation as well as the impact of quantum fluctuations introduced through the current density two-point correlation function are presented for two relevant electric field-to-mass parameter values $qE/m^2$, indicative of the strength of the backreaction process. Previous efforts utilized approximate solutions to the linear response equation that were expected to be valid for early times. A comparative analysis is given between the exact and approximate solutions in order to validate this conjecture.
    \end{abstract}

    \date{\today}
    \maketitle

\newpage

\section{Introduction}

    The semiclassical approach, which couples a quantized matter field to a classical background, has been applied across a wide range of scenarios~\cite{schwinger,parker,hawking,parkertoms,birrelldavies}. However, it is typically regarded as an approximate version of a fully quantized theory. Given that this approximation relies on the expectation values of objects such as the stress-energy tensor and the electric current that are constructed from quantum field operators, it is expected to break down when the associated quantum fluctuations of these objects are in some sense large.
    
    There are two distinct approaches to the semiclassical approximation. One is derived using a loop expansion of the effective action \cite{Schwartz}. In this approach, the semiclassical approximation fails when the quantum corrections become comparable to the classical background field, as this indicates higher-order terms in the loop expansion may become significant in that regime. In the second approach, the semiclassical approximation is derived using the large-$N$ expansion \cite{CooperHabib}, where $N$ identical quantum fields are coupled to the classical background field at leading order. Quantum corrections due to the background field first emerge at next-to-leading order. This approach allows for consistent solutions to the semiclassical backreaction equation when the quantum fields significantly influence the classical background field.

    There have been numerous efforts to develop a method to analyze the degree to which the semiclassical approximation is an accurate representation for a given physical model \cite{kuoford,phillipshu,wuford,AndersonParisMottola,AndersonParisSanders,PlaNewsomeAnderson}. Various correlation functions can be used to characterize quantum fluctuations, such as $\la T_{\mu \nu}(x) T_{\alpha \beta}(x') \ra$ in semiclassical gravity. However, it has been shown that complications arise when consideration is given to some of these which can manifest as state-dependent divergences in the coincident point limit~\cite{wuford}, varying results from different renormalization procedures~\cite{phillipshu}, and issues with covariance~\cite{AndersonParisMottola}.

    Within the framework of linear response theory~\cite{FetterWalecka,KapustaBellac,mottola}, an alternative method that does not suffer from these difficulties was developed by estimating the importance of certain quantum fluctuations using a criterion first formulated in~\cite{AndersonParisMottola} for semiclassical gravity. There it was applied to free scalar fields in flat spacetime evaluated in the Minkowski vacuum state. The criterion was also applied to the conformally invariant scalar field in the Bunch-Davies state in de Sitter space~\cite{AndersonParisMottola2}, with a modified version applied to preheating in models of chaotic inflation~\cite{AndersonParisSanders} and later to semiclassical electrodynamics~\cite{PlaNewsomeAnderson}.

    The criterion involves the stability of solutions to the linear response equation, which can be obtained by perturbing the semiclassical backreaction equation about a background field solution. In general, the linear response equation obtained in this way is an integro-differential equation which involves integration over the retarded two-point correlation function, thereby rendering the evolution of perturbations as manifestly causal. It can be shown this particular two-point function avoids the technical issues previously described, and therefore it is expected that any instability introduced through its presence in the linear response equation can be taken as one measure of the strength of quantum fluctuations associated with the quantum source term in the semiclassical backreaction equation. The criterion used here and in~\cite{PlaNewsomeAnderson} for the validity of the semiclassical approximation is violated if any linearized, gauge-invariant quantity constructed from solutions to the linear response equation, with finite nonsingular initial data, grows rapidly over some period of time. If this occurs, then quantum fluctuations are significant and the semiclassical approximation breaks down. Note that satisfaction of the criterion is a necessary but not sufficient condition. The linear response criterion provides a natural and well-defined way for this two-point correlation function to enter into the determination of the validity of the semiclassical approximation.

    In what follows, a semiclassical electrodynamics model in 1+1 dimensions is investigated where a quantized spin $\sfrac{1}{2}$ field evolves in the presence of a classical, spatially homogeneous electric field background which is generated by an external source. The semiclassical Maxwell field equations take the form
\be
    \partial_\mu F^{\mu \nu} = J^\nu_C + \bra{0_A} J^\nu_Q \ket{0_A} \quad , \label{semiclassMax}
\ee
    and are considered to replace the fully quantized theory. Here $J^\mu_C$ and $\la J^\mu_Q \ra$ represent classical and quantum source terms respectively, and the gauge field $A_\mu$ is taken to be a purely classical quantity upon which the modes defining the vacuum state $\ket{0_A}$ implicitly depend.

    The semiclassical backreaction equation has been solved for massive scalar and spin $\sfrac{1}{2}$ fields coupled to an electric field in 1+1 \cite{PlaNewsomeAnderson,Kluger91,Kluger92,Kluger93} and 3+1 \cite{Kluger93, Tanji, stat-FT}dimensions. The electric field
    was assumed to be homogeneous in space, but allowed to vary in time. In \cite{PlaNewsomeAnderson}, three classical current profiles were studied which generated electric fields with well-defined initial conditions. The first was a current that is proportional to a delta function potential, yielding an electric field that achieves a nonzero value instantaneously. The second involved the sudden turn-on of the classical current, but a gradual turn-on of the corresponding electric field which asymptotically approaches some constant value. The other is the well-known Sauter pulse generated by a current that has the form of a smooth pulse which only yields a significant classical electric field for a finite period of time. 

    From Schwinger's \cite{schwinger} original pair production rate $\Gamma \sim q^2 E^2 e^{-\pi m^2 / q E}$, it can be observed that the pair production intensity will be exponentially suppressed unless $qE \gtrsim m^2$. A critical scale can therefore be defined as $qE/m^2 \sim 1$, above which significant particle production is expected to occur. Experimental detection of the Schwinger effect via electron-positron pair production, which at present has yet to be observed, requires an electric field strength of order $E \sim 10^{18}$ V/m. A model akin to the Sauter pulse is a likely candidate for a background profile necessary to detect Schwinger pair production, see e.g. \cite{Kohlfurst} and references therein.

    The validity of the semiclassical approximation for quantum electrodynamics in 1+1 dimensions was investigated in~\cite{PlaNewsomeAnderson} by analyzing homogeneous solutions to the spin $\sfrac{1}{2}$ field linear response equation for classical sources which generate either an asymptotically constant or Sauter pulse electric field profile. There, explicit forms of the linear response equation were derived for both massive complex scalar and spin $\sfrac{1}{2}$ fields. A method of approximating the homogeneous solutions to the linear response equation for semiclassical electrodynamics was utilized. It was also used to investigate the validity of the semiclassical approximation during the preheating phase of chaotic inflation \cite{AndersonParisSanders}. The method involves computing the difference between two solutions to the semiclassical backreaction equation which have similar initial conditions, and was conjectured to be valid at early times. That conjecture is tested here by comparing the difference between two solutions to the semiclassical backreaction equations with the corresponding solution to the linear response equation.

    The criterion used here for the validity of the semiclassical approximation in electrodynamics is based upon the fact that the linear response equation depends on the retarded two-point correlation function $\la [ J_Q(t,x) , J_Q(t',x') ] \ra$ for the spin $\sfrac{1}{2}$ current density $J_Q$. It is expected that if quantum fluctuations associated with this correlation function are significant, then its impact will cause solutions to the linear response equation to grow significantly in time. To investigate this, a detailed analysis of the current density two-point function is given for the asymptotically constant electric field profile.
       
    For the numerical results, three values of the relative scale $qE_0/m^2$ are considered. They are the critical case for pair production $qE_0/m^2 =1$, a case where either the maximum classical electric field or the charge to mass ratio is relatively large, $qE_0/m^2 =10^3$, and for completeness, the limit in which the mass vanishes for fixed charge and maximum value of the classical electric field, $qE_0/m^2\to \infty$. In the massless case, an analytic solution to the semiclasscial backreaction equations was obtained in~\cite{PlaNewsomeAnderson} for the asymptotically constant profile. The corresponding solution to the linear response equation is given here.

    In Sec.~II, a review of the quantization for a spin $\sfrac{1}{2}$ field coupled to a classical source is presented. In Sec.~III, the semiclassical backreaction equation with both classical and renormalized quantum source terms is discussed. Details of the linear response formalism applied to the case of a spin $\sfrac{1}{2}$ field for both classical background profiles are presented in Sec.~IV. In Sec.~V, numerical results for solutions to the linear response equation, a comparative analysis between the exact and approximate linear response equation solutions used in \cite{PlaNewsomeAnderson}, and the behavior of the current density two-point function are included for both classical profiles. In Sec.~VI, a discussion of the results is presented. Appendix A contains a description of the numerical methods used to solve the linear response equation. Appendix B contains a comparison of the characteristic polarization tensor used in this paper for the linear response analysis and that used in a different approach involving perturbative quantum electrodynamics on a nontrivial background~\cite{Gitman,Fradkin,Shvartsman,Soviet}.

\section{Spin $\sfrac{1}{2}$ Field Quantization with a Classical Source}

    The action representing a free spin $\sfrac{1}{2}$ field $\psi$ coupled to a background electric field in two dimensions is
\be
    S[A_\mu, \Bar{\psi}, \psi] = \int d^{2}x \, \bigg[-\frac{1}{4}F_{\mu \nu}F^{\mu \nu} + A_\mu J_C^\mu +i\Bar{\psi}\gamma^{\mu}D_{\mu}\psi-m\Bar{\psi}\psi\bigg] \label{SS} \quad .
\ee
    Here $F_{\mu \nu}=\partial_\mu A_\nu - \partial_{\nu}A_{\mu}$ is the electromagnetic field strength tensor with gauge field $A_\mu$, the term $J^\mu_C$ is a classical and conserved external source, and $D_{\mu}=\partial_{\mu}-i q A_{\mu}$ is the gauge covariant derivative with charge $q$. The adjoint of the spin $\sfrac{1}{2}$ field is $\Bar{\psi}=\psi^{\dagger}\gamma^0$, with $m$ the mass of the field. The Dirac matrices $\gamma^{\mu}$ satisfy $\{\gamma^\mu , \gamma^\nu\}=-2 \eta^{\mu \nu}$. The metric signature is chosen to be $(-,+)$ with the unit convention $\hbar =c=1$.

    Variation of \eqref{SS} with respect to the vector potential yields the general form of Maxwell's equations with sources
\be
    -\Box A^\mu + \partial^\mu \partial_\nu A^\nu = J^\mu_C + J^\mu_Q \quad . \label{Max}
\ee
    The conserved quantum current density is
\be
    J^{\mu}_Q = q \, \Bar{\psi}(t,x) \, \gamma^{\mu} \, \psi(t,x) \label{J2} \quad .
\ee
    Variation of $\eqref{SS}$ with respect to $\Bar{\psi}$ yields the Dirac equation
\be
    \big[i \, \gamma^{\mu}D_{\mu}-m\, \big]\psi(t,x) = 0 \label{mode2} \quad .
\ee
    In what follows, the Lorentz gauge $\partial_\mu A^\mu=0$ is chosen and the vector potential is fixed to have the form
\be
    A^\mu = (0,A(t)) \quad , \label{gauge} 
\ee
    which yields $F_{01}=\partial_0 A_1=\dot{A}=-E$. Then \eqref{mode2} becomes
\be
    \bigg[i \, \gamma^{t}\partial_{t}+i \, \gamma^{x}\partial_{x}+q \, \gamma^{x}A(t)-m \bigg]\psi(t,x) =0 \quad . \label{modefermi}
\ee
    The spin $\sfrac{1}{2}$ field can be expanded in terms of a complete set of basis mode functions as
\be
    \psi(t,x) = \int_{-\infty}^\infty dk \, \bigg[B_{k}u_{k}(t,x)+D_{k}^{\dagger}v_{k}(t,x)\bigg] \quad , \label{psi1}
\ee
    where $B_{k},B_{k}^{\dagger},D_{k}$, and $D_{k}^{\dagger}$ are the creation and annihilation operators obeying the canonical anticommutation relations $\{B_{k}, B^{\dagger}_{k'}\} = \{D_{k}, D^{\dagger}_{k'}\} = \delta(k-k')$. Due to the assumption of spatial homogeneity and utilizing the Weyl representation of the Dirac matrices
\be 
    \gamma^{t} =
    \begin{bmatrix}
        0 && 1  \\
        1 && 0  \\
    \end{bmatrix} \quad , \quad
    \gamma^{x} =
    \begin{bmatrix}
        0 && 1  \\
        -1 && 0  \\
    \end{bmatrix} \quad , \quad
    \gamma^{5} =\gamma^{t}\gamma^{x}=
    \begin{bmatrix}
        -1 && 0  \\
        0 && 1 \\
    \end{bmatrix}
\quad , \label{weylM}
\ee
    one can construct two independent spinor solutions \cite{h1h2modes} as 
\be
    u_{k}(t,x)=\frac{e^{ikx}}{\sqrt{2\pi}}
    \begin{bmatrix}
    h_{k}^{I}(t) \\
    -h_{k}^{II}(t)
    \end{bmatrix} \quad , \quad
    v_{k}(t,x)=\frac{e^{-ikx}}{\sqrt{2\pi}}
    \begin{bmatrix}
    h_{-k}^{II *}(t) \\
    h_{-k}^{I *}(t)
    \end{bmatrix} \quad . \label{v}
\ee
    Substituting $\eqref{v}$ into $\eqref{modefermi}$ one finds $h^{(I,II)}_k$ satisfy the following equations
\bes 
    \be
        \frac{d}{dt}h_{k}^{I}(t)-i\bigg[k-qA(t)\bigg]h_{k}^{I}(t)-i\,m\,h_{k}^{II}(t) = 0 \quad , \label{modeh1}
    \ee
    \be
        \frac{d}{dt}h_{k}^{II}(t)+i\bigg[k-qA(t)\bigg]h_{k}^{II}(t)-i\,m\,h_{k}^{I}(t) = 0 \quad . \label{modeh2}
    \ee \label{h1h2modes}
\ees
    The normalization condition $|h_k^{I}|^{2}+|h_k^{II}|^{2}=1$ ensures the anticommutation relations are satisfied.

    It is useful to mention two distinct limits of the solutions to $\eqref{h1h2modes}$. The first is the limit in which the electric field and the vector potential vanish. This is relevant for times $ t \le 0$ for the asymptotically constant profile and for the limit $t \to - \infty$ for the Sauter pulse. These solutions are
\bes
    \begin{flalign}
        h_k^{I}(t) &=  \sqrt{\frac{\omega - k}{2 \, \omega}} e^{-i \omega t} \quad , \\
        h_k^{II}(t) &= - \sqrt{\frac{\omega + k}{2 \, \omega}} e^{-i \omega t} \quad ,
    \end{flalign}
    \label{h1h2vacuum}
\ees
    where $\omega^2=k^2+m^2$. Note that since the classical current is initially zero, and gives rise to an electric field that is initially zero as well, there is no ambiguity in the choice of vacuum state. The second is the massless limit in which the mode equations $\eqref{h1h2modes}$ decouple and have the general solutions \cite{PlaNewsomeAnderson}
\bes
    \begin{flalign}
        h^{I}_k(t) &=  \theta(- k) e^{ i\int_{t_0}^t\left[ k-q A(t') \right]dt'} \quad , \\
        h^{II}_k(t) &= - \theta( k) e^{- i\int_{t_0}^t\left[ k-q A(t') \right]dt'} \quad .
    \end{flalign}
    \label{h1h2massless}
\ees
    Here $\theta(x)$ is the Heaviside step function and $t_0$ indicates the initial time at which the electric field turns on. This solution is consistent with the vacuum state when the background source is shut off.

\section{The Semiclassical Backreaction Equation}

    The time evolution of the electric field, which arises from a classical source and is subsequently modified through quantum effects, is governed by the semiclassical backreaction equation. This is obtained by the replacement $J_Q^\mu \to \la J_Q^\mu \ra$ in $\eqref{Max}$. In the Lorentz gauge with choice $\eqref{gauge}$ and for $\mu=1$, this becomes
\be
    \frac{d^{2}}{dt^{2}} A(t) =-\frac{d }{d t}E(t)=J_{C}(t) + \langle J_{Q}(t) \rangle_\textnormal{ren} \quad , \label{sb}
\ee
    where $J_C \equiv J_C^1$ and $\la J_Q \ra_\mathrm{ren} \equiv \la J_Q^1 \ra_\mathrm{ren}$. The $\mu=0$ component for either source corresponds to the induced electric charge and is identically zero, i.e. no net charge is generated.

    Two separate classical background profiles are chosen to couple to the spin $\sfrac{1}{2}$ field. The asymptotically constant profile has a current density source and electric field of the form
\bes
    \be
        J_{C}(t) = -\frac{ q E_{0} }{(1+qt)^{2}} \label{Jasymp} \quad ,
    \ee
    \be
        E_C(t) = -\int_{0}^t J_C(t') \, dt' =E_{0}\left(\frac{qt}{1+qt}\right) \quad , \label{Easymp}
    \ee \label{asymp}
\ees
    for $t\geq 0$. The second choice of profile is the Sauter pulse, with current density source and electric field given by
\bes
    \be
        J_{C}(t) = -2qE_{0} \, \textnormal{sech}^2(qt)\textnormal{tanh}(qt) \label{JSauter} \quad ,
    \ee
    \be
        E_{C}(t)=E_{0} \, \textnormal{sech}^{2}(qt) \label{ESauter} \quad ,
    \ee \label{Sauter}
\ees
    for $-\infty < t < \infty$.
    
    The renormalized expression for the current $\la J_{Q} \ra$ can be found by evaluating \eqref{J2} in the vacuum state and is given by \cite{PlaNewsomeAnderson,h1h2modes}
\be
    \la J_{Q}(t) \ra_{\textnormal{ren}} =-\frac{q^2}{\pi}A(t) +  \frac{q}{2\pi}\int_{-\infty}^{\infty} dk \, \bigg[ |h_{k}^{I}(t)|^{2}-|h_{k}^{II}(t)|^{2}+\frac{k}{\omega}\bigg] \label{Jrenorm}
\quad .
\ee
    Here the procedure of adiabatic regularization has been used to eliminate the ultraviolet divergence \cite{PlaNewsomeAnderson}.
    
    As particle production occurs, the electric field originating from $J_C$ will accelerate the created particles. The counter-electric field produced by the current density $\la J_Q \ra_\mathrm{ren}$ will initially be in opposition to, and therefore begin to cancel, the original background electric field. This field is completely canceled after some characteristic period of time depending upon the relative size of $qE/m^2$. The result is an electric field with an opposite orientation compared with the original background field due to the continued motion of the spin $\sfrac{1}{2}$ particles. If particle interactions are ignored, the process will continue indefinitely, with the particles undergoing plasma oscillations and the total electric field oscillating in time.

\subsection{Massless Limit}
    
    In the massless limit, substitution of~\eqref{h1h2massless} into~\eqref{Jrenorm} gives
\be
    \lim_{m \, \rightarrow \, 0} \langle J_Q(t) \rangle_\textnormal{ren} = -\frac{q^2}{\pi}A(t) \label{masslessJq} \quad ,
\ee
    and therefore $\eqref{sb}$ becomes
\be
    \frac{d^2}{dt^2} A(t) + \frac{q^2}{\pi} A(t) = J_C(t) \quad , \label{Amassless}
\ee
    which is the equation for a harmonic oscillator with frequency $|q|/\sqrt{\pi}$ and source $J_C$. 
    
    For the classical source $\eqref{Jasymp}$ with initial conditions $A(0)=0$ and $E(0)=0$, the solution to $\eqref{Amassless}$, is given by~\cite{PlaNewsomeAnderson}
\begin{flalign}
    A(t) = -\frac{E_0}{q} &\bigg\{ \cos{\left( \frac{1+qt}{\sqrt{\pi}} \right)} \left[ \mathrm{ci}\left( \frac{1}{\sqrt{\pi}} \right) - \mathrm{ci}\left(\frac{1+qt}{\sqrt{\pi}}\right) \right] \nonumber \\
    &+ \sin{\left( \frac{1+qt}{\sqrt{\pi}} \right)} \left[ \mathrm{si}\left( \frac{1}{\sqrt{\pi}} \right) - \mathrm{si}\left(\frac{1+qt}{\sqrt{\pi}}\right) \right] \nonumber \\
    &+\sqrt{\pi}\sin{\left( \frac{qt}{\sqrt{\pi}} \right)} \bigg\} \quad . \label{Anomass}
\end{flalign}
    The electric field is
\begin{flalign}
    E(t) = E_0 &\bigg\{\frac{1}{\sqrt{\pi}}\sin{\left( \frac{1+qt}{\sqrt{\pi}} \right)}\bigg[ \textnormal{ci}\left( \frac{1+qt}{\sqrt{\pi}} \right) - \textnormal{ci}\left( \frac{1}{\sqrt{\pi}}  \right) \bigg] \nonumber \\
    &+ \frac{1}{\sqrt{\pi}}\cos{\left( \frac{1+qt}{\sqrt{\pi}} \right)} \bigg[ \textnormal{si}\left( \frac{1}{\sqrt{\pi}} \right) - \textnormal{si}\left( \frac{1+qt}{\sqrt{\pi}} \right) \bigg] \nonumber \\
    &+ \cos{\left( \frac{qt}{\sqrt{\pi}} \right)} - \frac{1}{1+qt} \bigg\} \quad , \label{Enomass}
\end{flalign}
    where $\mathrm{ci}(x)=-\int_x^\infty \frac{\cos{(t)}}{t}dt$ and $\mathrm{si}(x)=-\int_0^x \frac{\sin{(t)}}{t}dt$ are the cosine and sine integral functions respectively. There is no convenient analytic form for solutions to $\eqref{Amassless}$ for the Sauter pulse classical source \eqref{JSauter}. However, it is expected that solutions will be characterized by similar harmonic behavior as is evidenced numerically in Sec.~\ref{Sec5a}.

\section{The Linear Response Equation}
\label{LREsection}
    Formally, the semiclassical linear response equation can be derived by taking the second variation of the effective action. However, for the 1+1 dimensional electrodynamics model considered here the linear response equation can more simply, but equivalently, be obtained by perturbing the semiclassical backreaction equation \eqref{sb} about a background solution such that $A \to A+\delta A$, resulting in 
\be
    \frac{d^{2}}{dt^{2}}\delta A(t)=-\frac{d}{d t}\delta  E(t) = \delta J_{C}(t)+\delta \langle J_{Q}(t)\rangle_\textnormal{ren} \quad . \label{LRE}
\ee
    The type of perturbation being considered is one that is driven by changes in the classical current density $J_C \to J_C + \delta J_C$. Thus, for $\eqref{Jasymp}$ the classical perturbation is
\be
    \delta J_C(t) = - \frac{q}{(1+qt)^2} \delta E_0 \quad , \label{deltaJclass}
\ee
    and for the Sauter pulse classical profile one has from \eqref{JSauter}
\be
    \delta J_C(t) = 2 q \, \sech^2{\left(q t\right)} \tanh{\left( q t \right)} \delta E_0  \quad . \label{deltaJclass2}
\ee
    A perturbation in the classical current density will necessarily induce a response from $\la J_Q \ra_\mathrm{ren}$ to this perturbation since the mode equation $\eqref{h1h2modes}$ depends on the background field $A$. The leading order contribution is \cite{PlaNewsomeAnderson}
\be
    \delta \la J_Q(t) \ra _\textnormal{ren} = -\frac{q^2}{\pi} \delta A(t) + i \int_{-\infty}^{t} dt' \int_{-\infty}^{\infty} dx' \, \la [J_Q(t,x) , J_Q(t',x')] \ra \, \delta A(t') \label{deltaJqrenorm}
\ee
    Excluding the first term on the right hand side, which originates from the adiabatic regularization of \eqref{Jrenorm}, the form \eqref{deltaJqrenorm} takes is a general feature of linear response theory. Here, $\delta A(t')$ acts as a source at a past time $t'$ which induces a change $\delta \la J_Q(t) \ra _\textnormal{ren}$ measured at the present time $t$. Therefore, the two-point function $\la [J_Q(t,x) , J_Q(t',x')] \ra$ can be associated with a generalized susceptibility \cite{FetterWalecka}.\footnote{See Appendix B for a discussion of the retarded two-point correlation function, its relationship to the polarization tensor, and how it compares with that of scattering theory in perturbative quantum electrodynamics on a nontrivial background.} Note that for the cases considered here, the vector potential and its first time derivative are initially zero. As a result, these perturbations do not cause a change in the vacuum state for the field.

    The retarded two-point function can be expressed in terms of a product of mode functions $\eqref{v}$ as
\be
    \la [J_Q(t,x) , J_Q(t',x')] \ra = \frac{q^2}{2\pi^2}\, i \int_{-\infty}^{\infty} dk \int_{-\infty}^{\infty} dk' \, e^{-i \left( k - k' \right)\left( x - x' \right)} \textnormal{Im}\bigg\{ f_{k , k'}(t) \, f^*_{k , k'}(t') \bigg\} \quad , \label{twopoint2}
\ee
    with
\be
    f_{k , k'}(t) \equiv h_{k'}^{I}(t)h_{k}^{II}(t)+h_{k}^{I}(t)h_{k'}^{II}(t) \quad .
\ee
    Subsequently, the spatial integral present in $\eqref{deltaJqrenorm}$ is \cite{PlaNewsomeAnderson}
\be
    -i \int_{-\infty}^{\infty} dx' \, \la [J_Q(t,x) , J_Q(t',x')] \ra = \frac{4q^2}{\pi} \int_{-\infty}^{\infty} dk \, \textnormal{Im}\left\{h_k^I(t)h_k^{II}(t)h_k^{I*}(t')h_k^{II*}(t')\right\} \quad . \label{twopoint1}
\ee
    Thus \eqref{LRE} with either perturbed classical source term $\eqref{deltaJclass}$ or $\eqref{deltaJclass2}$ takes the form
\begin{flalign}
    \frac{d^2}{dt^2}\delta A(t) = \delta J_C(t) - \frac{q^2}{\pi} \delta A(t) - \frac{4 q^2}{\pi} \int_{-\infty}^{t} dt' \int_{-\infty}^{\infty} dk \, \textnormal{Im}\left\{h_k^I(t)h_k^{II}(t)h_k^{I*}(t')h_k^{II*}(t')\right\} \delta A(t') \, \, . \label{fullLRE}
\end{flalign}

\subsection{Correlations Prior to Activation of the Background Field}

    It is interesting to note that the retarded current density two-point correlation function is nonzero even when the classical current is zero.  Substituting~\eqref{h1h2vacuum} into~\eqref{twopoint2} one finds
\be
    \la [J_Q(t,x) , J_Q(t',x')] \ra = -\frac{q^2}{8\pi^2} \, i \int_{-\infty}^\infty dk \int_{-\infty}^\infty \, dk' \, \chi_{k,k'}^2 \, e^{-i(k-k')(x-x')} \sin \big\{ (\omega_k+\omega_{k'})(t-t') \big\} \quad , \label{adiabaticJJ}
\ee
    where
\be
    \chi_{k,k'} = \frac{1}{\sqrt{\omega_k \omega_{k'}}} \bigg[ \sqrt{\big(\omega_k + k\big)\big(\omega_{k'}-k'\big)} + \sqrt{\big(\omega_k - k\big)\big(\omega_{k'}+k'\big)} \bigg] \quad . \label{adiabaticchi}
\ee
    In $\eqref{adiabaticJJ}$ and $\eqref{adiabaticchi}$ the notation has been slightly modified to distinguish $\omega^2_k \equiv k^2+m^2$ and $\omega^2_{k'}\equiv k'^{\,2}+m^2$. It follows that $\eqref{twopoint1}$ reduces to
\be
    - i \int_{-\infty}^{\infty} dx' \, \la [J_Q(t,x) , J_Q(t',x')] \ra = \frac{q^2 m^2}{\pi} \int_{-\infty}^\infty dk \, \bigg[ \frac{\sin\big\{ 2 \omega \big(t'-t\big) \big\}}{\omega^2} \bigg] \quad . \label{JJNoField}
\ee
    Since no particles are present prior to activation of the classical current $J_C$, the fact that the two-point function is nonzero indicates the existence of vacuum polarization effects in the absence of a background electric field.

\subsection{Massless Limit}

    When the time-dependent mode functions $h^{(I,II)}_k$ take the form $\eqref{h1h2massless}$, the current density two-point function $\eqref{twopoint2}$ vanishes. It then follows from $\eqref{masslessJq}$ that $\eqref{deltaJqrenorm}$ simplifies to
\be
    \lim_{m\rightarrow 0} \delta \la J_Q(t) \ra_\textnormal{ren} = -\frac{q^2}{\pi} \delta A(t) \quad . \label{masslesscurrent}
\ee
    Thus the linear response equation \eqref{LRE} with $\eqref{masslesscurrent}$ becomes
\be
    \frac{d^2}{dt^2}\delta A(t) + \frac{q^2}{\pi}\delta A(t) = \delta J_C(t) \quad , \label{LREmassless}
\ee
    which is of similar harmonic character as $\eqref{Amassless}$ with equivalent frequency. The form of $\eqref{LREmassless}$ guarantees for the massless limit that perturbations in the background field remain bounded with fixed amplitude for both forms of $\delta J_C$ in \eqref{deltaJclass} and \eqref{deltaJclass2}. For the case of $\eqref{deltaJclass}$, the solution to $\eqref{LREmassless}$ with initial conditions $\delta A(0)=0$ and $\delta E(0)=0$ yields the same result as in $\eqref{Anomass}$ and $\eqref{Enomass}$, but with the replacement $E_0 \to \delta E_0$.

    It is interesting to note that given the limit $\eqref{masslesscurrent}$, the non-local term in $\eqref{deltaJqrenorm}$ carries the additional interpretation of being a measure of the extent to which $\delta \la J_Q \ra_\mathrm{ren}$ differs from that of its massless counterpart, since
\be
    \delta \la J_Q(t) \ra _\textnormal{ren} = \lim_{m\rightarrow 0} \delta \la J_Q(t) \ra_\textnormal{ren} + i \int_{-\infty}^{t} dt' \int_{-\infty}^{\infty} dx' \, \la [J_Q(t,x) , J_Q(t',x')] \ra \, \delta A(t') \quad . \label{masslesscomp}
\ee

\section{Numerical Results}

    The relative scale between the electric field strength $E_0$ and the mass $m$ associated with spin $\sfrac{1}{2}$ particles is characterized by the quantity $qE_0/m^2$. For all numerical results presented, the cases $qE_0/m^2 = 1$ and $qE_0/m^2 = 10^3$ are considered, the former being identified with the critical threshold for Schwinger pair production in the case of a constant electric field. A larger value of $qE_0/m^2$ corresponds to an electric field which has a higher energy density to fuel the pair production process, generating on average larger backreaction effects resulting from a more significant quantum current density $\langle J_Q \rangle$. In the large mass limit $qE/m^2 \to 0$, the electric field will not supply sufficient energy to create particles, so one expects that $\la J_Q \ra_\mathrm{ren} \to 0$ and $E \to E_C$. As highlighted in~\cite{PlaNewsomeAnderson}, this outcome is consistent with the decoupling theorem in perturbative quantum field theory~\cite{AppelquistCarazzone}, which states that heavy masses decouple in the low-energy description of the theory. Here, in the limit $m^2 \to \infty$ while keeping $E_0$ fixed, the theory simplifies to classical electrodynamics.

\subsection{Backreaction and Linear Response Equation Solutions}
\label{Sec5a}

\begin{figure}[h]
    \centering
    \includegraphics[scale=0.425]{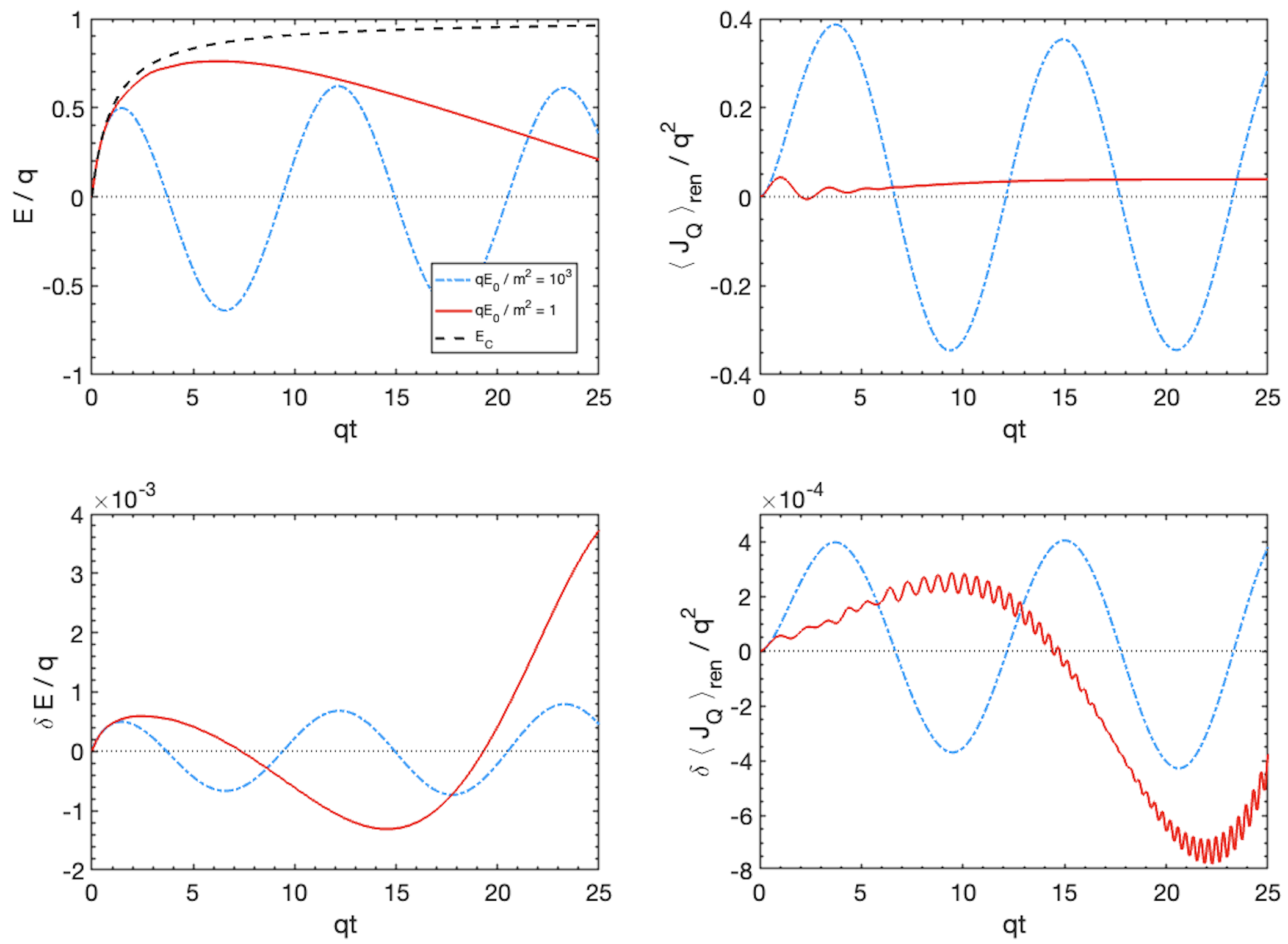} 
    \caption{The classical electric field $E/q$ including backreaction effects (top-left), the associated spin $\sfrac{1}{2}$ current density $\la J_Q \ra_\mathrm{ren}/q^2$ (top-right), the linear response solution $\delta E / q$ (bottom-left), and the associated perturbation of the current density $\delta \la J_Q \ra_\mathrm{ren}/q^2$ (bottom-right), are all shown as a function of time $qt$ for the classical sources $\eqref{Jasymp}$ and $\eqref{deltaJclass}$. The characteristic cases $qE_0/m^2=1$ and $qE_0/m^2=10^3$ are shown for all plots, with the classical electric field solution also included. For both cases, $E_0/q=1$ so that $m^2/q^2=1$ for the top row plots and $m^2/q^2=10^{-3}$ for the bottom row plots. Included for comparison, the dotted horizontal curve indicates when the corresponding quantity being plotted is zero.}
    \label{Fig1}
\end{figure}

\begin{figure}[h]
    \centering
    \includegraphics[scale=0.425]{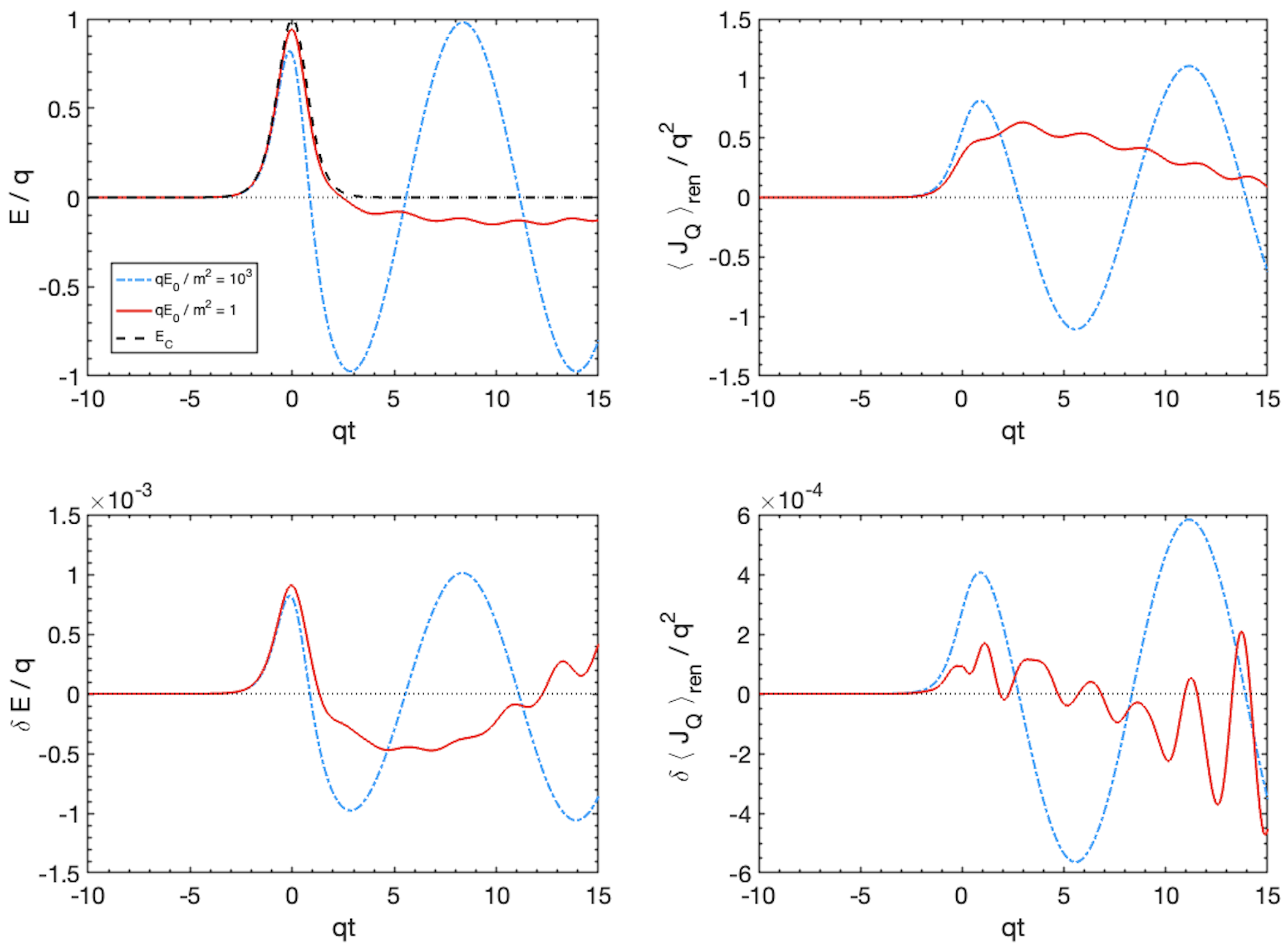} 
    \caption{The electric field $E/q$ including backreaction effects (top-left), the associated spin $\sfrac{1}{2}$ current density $\la J_Q \ra_\mathrm{ren}/q^2$ (top-right), the linear response solution $\delta E / q$ (bottom-left), and the associated perturbation of the current density $\delta \la J_Q \ra_\mathrm{ren}/q^2$ (bottom-right), are all shown as a function of time $qt$ for the classical sources $\eqref{JSauter}$ and $\eqref{deltaJclass2}$. The characteristic cases $qE_0/m^2=1$ and $qE_0/m^2=10^3$ are shown for all plots, with the classical electric field solution also included. For both cases, $E_0/q=1$ so that $m^2/q^2=1$ for the top row plots and $m^2/q^2=10^{-3}$ for the bottom row plots. Included for comparison, the dotted horizontal curve indicates when the corresponding quantity being plotted is zero.}
    \label{Fig2}
\end{figure}

    For the classical source terms $\eqref{Jasymp}$ and $\eqref{JSauter}$ associated with the asymptotically constant and Sauter pulse profiles, both the backreaction process, represented through the electric field $E$, the renormalized current density $\langle J_Q \rangle_\mathrm{ren}$, and their associated linear response to perturbations, $\delta E$ and $\delta \la J_Q \ra_\mathrm{ren}$, are shown in Fig.~\ref{Fig1} and Fig.~\ref{Fig2} respectively. The results for $E$ and $\langle J_Q \rangle_\mathrm{ren}$ were previously shown in~\cite{PlaNewsomeAnderson}. Note that in this subsection the focus is on the behaviors of solutions to the linear response equation. In Sec.~VB, a discussion is given between these solutions and the validity of the semiclassical approximation.

    For $qE_0/m^2=1$, the deviation of the electric field from the classical solution for the asymptotically constant profile increases monotonically over the times considered. In contrast for the Sauter pulse profile, after the maximum of the pulse occurs at time $qt = 0$, the electric field undergoes long period oscillations which are modulated by shorter-period, smaller-amplitude oscillations. For $qE_0/m^2=10^3$, backreaction effects are significantly stronger. In all cases plasma oscillations occur at sufficiently late times. The strongest backreaction effects occur in the massless limit, $q E_0/m^2 \to \infty$, resulting in pure harmonic behavior for the electric field.
    
    The current density $\la J_Q \ra_\mathrm{ren}$ in the case $qE_0/m^2=1$ for the asymptotically constant profile exhibits small oscillations at early times which decay away, likely having their origin in the sudden activation of the background current $J_C$. At late times, large timescale oscillations are present. For the Sauter pulse profile, once the pulse occurs at $qt=0$ there is an initial increase in the current density which begins to dampen soon after. The higher frequency modulations present are similar in nature to those seen for very early times in the case of the asymptotically constant profile. However, for the Sauter pulse these modulations do not damp away. The amount of damping appears to be correlated with the size of the classical current, which after the Sauter pulse is very small. For $qE_0/m^2=10^3$, the current density for both profiles undergoes a more rapid increase with larger amplitude, indicative of an increase in pair production events, and begins to oscillate with an approximately constant frequency.

    The linear response solutions $\delta E$ to $\eqref{fullLRE}$ for the asymptotically constant profile with $qE_0/m^2=1$ undergo oscillatory behavior with an amplitude that grows in time, indicative of an instability. For the Sauter pulse when  $qE_0/m^2=1$, after the maximum of the pulse occurs at time $qt = 0$, the solution $\delta E$ undergoes long period oscillations which are modulated by shorter-period, smaller-amplitude oscillations which grow in time. For the case of $qE_0/m^2=10^3$, both solutions are characterized by approximate simple harmonic behavior with a frequency similar to that of the electric field.

    The associated current density perturbation $\delta \la J_Q \ra_\mathrm{ren}$ in the case of $qE_0/m^2=1$ for the asymptotically constant profile exhibits relatively large amplitude, long time-scale oscillations as well as smaller amplitude, higher frequency modulations. The Sauter pulse current density perturbation for $qE_0/m^2=1$ after $qt=0$ initially exhibits sporadic oscillatory behavior before settling down to constant frequency oscillations which grow in amplitude, generating the modulations seen at late times for $\delta E$. Both profiles for the case of $qE_0/m^2=10^3$ exhibit approximate simple harmonic behavior for $\delta \langle J_Q \rangle_\mathrm{ren}$, growing to a relatively constant amplitude with oscillations of constant frequency similar to that of $\la J_Q \ra_\mathrm{ren}$.

\subsection{Exact vs. Approximate Solutions to the Linear Response Equation}

    There are various way in which one can perturb the semiclassical backreaction equation, each yielding a modified solution $E_2$ that differs from the original solution $E_1$. The expansion of the electric field $E_2$ in terms of the unperturbed field $E_1$ is of the form $E_2=E_1+\delta E + \mathcal{O}(\delta E^2)$, where $\delta E$ is a solution to the linear response equation \eqref{LRE}. For two solutions $E_1$ and $E_2$ to the backreaction equation $\eqref{sb}$, whose initial conditions differ by a sufficiently small amount, one can construct $\Delta E \equiv E_2-E_1$ such that to linear order the perturbations can be approximated as $\Delta E \approx \delta E$. One would expect such an approximation to hold at early times prior to significant particle production. Here, two solutions are considered with slightly different values of $E_0$ for either the asymptotically constant classical profile \eqref{Easymp} or the Sauter pulse \eqref{ESauter}. For both profiles, $E_0$ is the maximum value the classical electric field will have.

   The difference between two solutions to the backreaction equation $\eqref{sb}$ satisfies
\be
    -\frac{d}{dt}\Delta E(t) = \Delta J_C(t) + \Delta \la J_Q(t) \ra_\textnormal{ren} \label{approxLRE} \quad ,
\ee
    with
\bes
    \be
        \Delta J_C(t) = J_{C,2}(t) - J_{C,1}(t) \quad ,
    \ee
    \be
        \Delta \la J_Q(t) \ra_\textnormal{ren} = \la J_{Q,2}(t) \ra_\textnormal{ren} - \la J_{Q,1}(t) \ra_\textnormal{ren} \quad .
    \ee
\ees
    In order for $\Delta E \approx \delta E$, it is clear that $\Delta \la J_Q \ra_\textnormal{ren} \approx \delta \la J_Q \ra_\textnormal{ren}$ must hold, since one can set $\Delta J_C = \delta J_C$ for all times.

    To analyze the behaviors of solutions to $\eqref{LRE}$ and compare them with $\eqref{approxLRE}$, it is useful to isolate the quantum contribution to the electric field $E_Q$ by subtracting from the exact electric field $E$ the corresponding solution to the classical equation $E_C$ as
\be
    E_Q(t) \equiv E(t)  - E_C(t) \quad .  \label{Ec-def}
\ee
    From the structure of $\eqref{LRE}$ and $\eqref{approxLRE}$, the quantum contributions to the exact and approximate solutions to the linear response equation $\delta E$ and $\Delta E$ can be similarly isolated, i.e. $\delta E_Q \equiv \delta E - \delta E_C$ and $\Delta E_Q \equiv E - E_C$. The criterion for the validity of the semiclassical approximation can therefore be modified to state that if quantities constructed from either $\delta E_Q$ or $\Delta E_Q$ grow significantly during some period of time then the semiclassical approximation is considered to be invalid~\cite{PlaNewsomeAnderson}.

    In order to provide a meaningful description of the growth in time for $\delta E_Q$, a useful quantity to consider is the relative difference between $\delta E_Q$ or $\Delta E_Q$ and the quantum contribution to the associated background field $E_Q$, formulated as\footnote{For ease of comparison with the solutions to the linear response equation, we use a slightly different definition of $R_Q$ than was used in~\cite{PlaNewsomeAnderson}.}
\bes
    \be
        R_Q(t) \equiv \frac{|\Delta E_Q(t)|}{|E_{Q,1}(t)|} \label{RqDeltaE} \quad ,
    \ee
    \be
        \Bar{R}_Q(t) \equiv \frac{|\delta E_Q(t)|}{|E_Q(t)|} \label{RqdeltaE} \quad .
    \ee \label{arsss}
\ees  
    The degree to which $\Delta E_Q \approx \delta E_Q$ will be characterized in a scale invariant way by the degree to which $R_Q \approx \Bar{R}_Q$. In~\cite{PlaNewsomeAnderson}, $R_Q$ was used to characterize the growth of the approximate solutions to the linear response equation. For the classical solutions one can define in the same way as above
\be
    \bar{R}_C \equiv \frac{|\delta E_C(t)|}{E_C(t)}
\ee

\begin{figure}[h]
    \centering
    \includegraphics[scale=0.455]{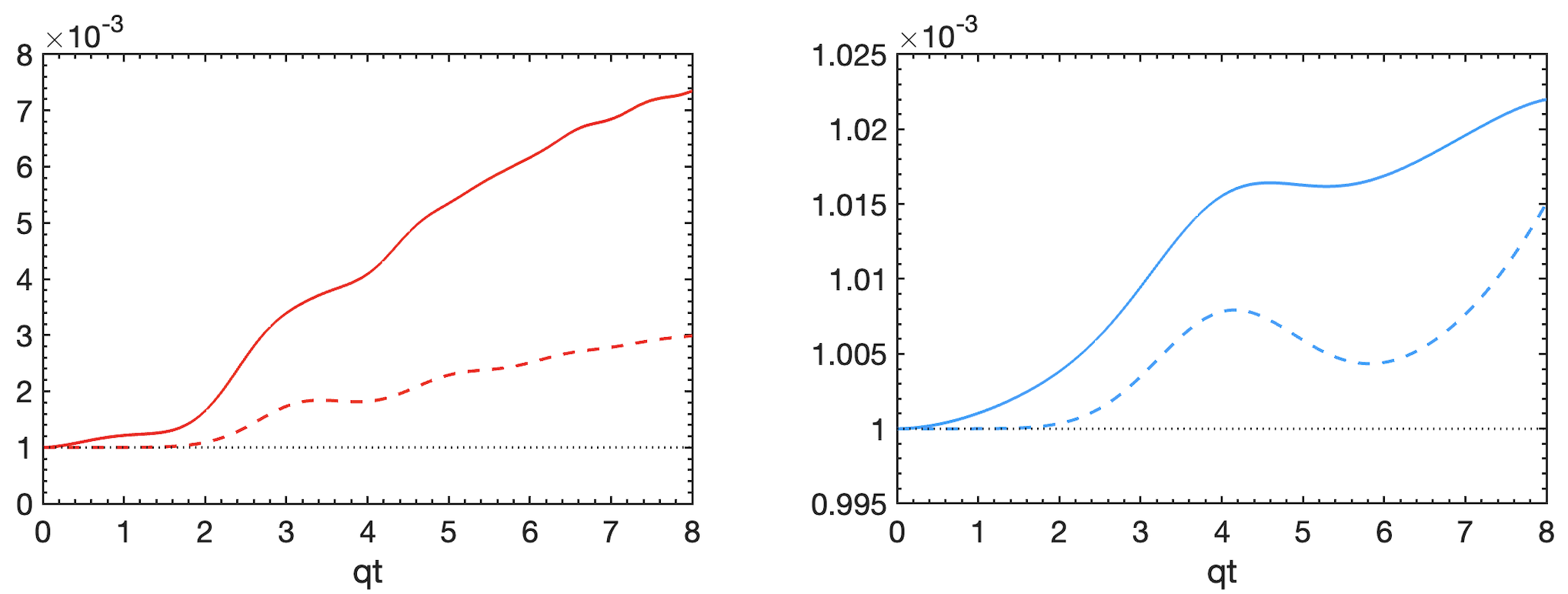} 
    \caption{The early time behavior is shown for the quantities $R_Q$ (dashed curve) and $\Bar{R}_Q$ (solid curve) for cases $qE_0/m^2=1$ (left) and $qE_0/m^2=10^3$ (right), and for the classical current $\eqref{Jasymp}$. The quantity $\bar{R}_C$ (black-dotted curve) is included for comparison. For both cases, $E_0/q=1$ so that $m^2/q^2=1$ for the left plot and $m^2/q^2=10^{-3}$ for the right plot.}
    \label{Fig3}
\end{figure}

\begin{figure}[h]
    \centering
    \includegraphics[scale=0.45]{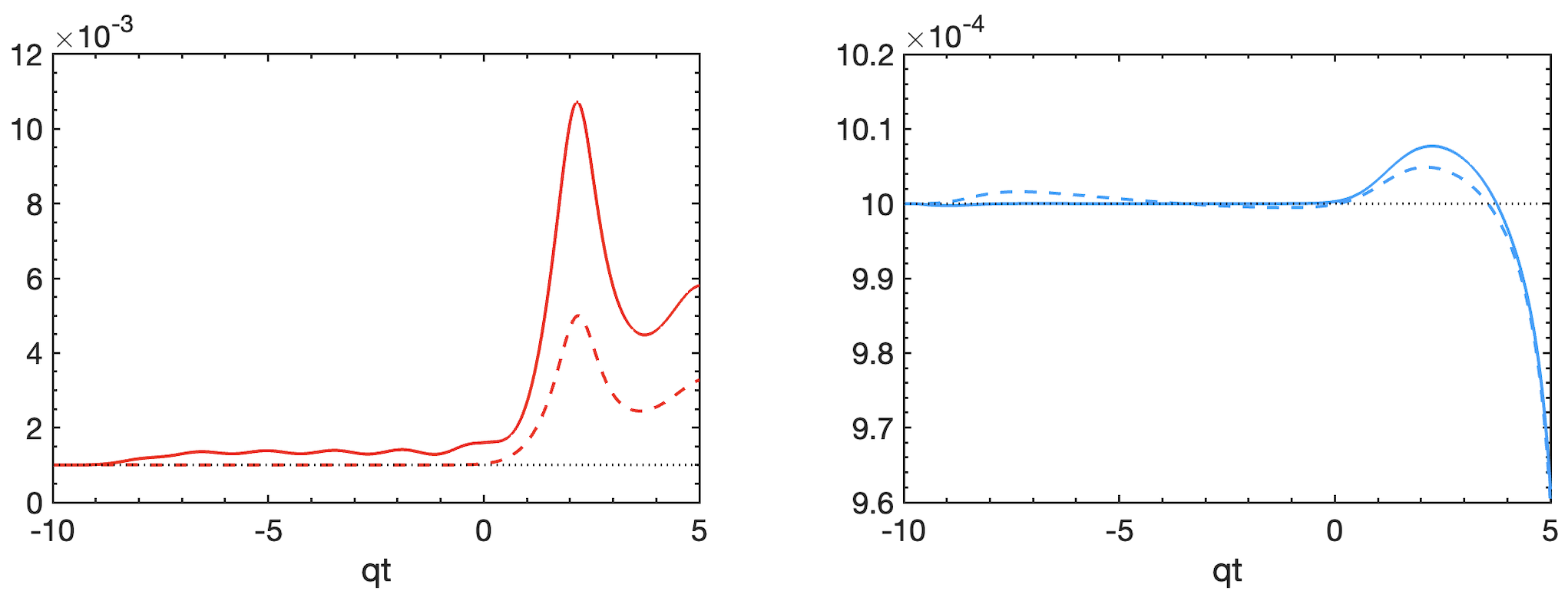} 
    \caption{The early time behavior is shown for the quantities $R_Q$ (dashed curve) and $\Bar{R}_Q$ (solid curve) for cases $qE_0/m^2=1$ (left) and $qE_0/m^2=10^3$ (right), and for the classical current $\eqref{JSauter}$. The quantity $\bar{R}_C$ (black-dotted curve) is included for comparison. For both cases, $E_0/q=1$ so that $m^2/q^2=1$ for the left plot and $m^2/q^2=10^{-3}$ for the right plot.}
    \label{Fig4}
\end{figure}

    The early time behaviors of both $R_Q$ and $\Bar{R}_Q$ are shown in Fig.~\ref{Fig3} for the asymptotically constant background profile and in Fig.~\ref{Fig4} for the Sauter pulse background profile. For both profiles, the perturbed electric field present in $\eqref{deltaJclass}$ and $\eqref{deltaJclass2}$ was set to $\delta E_0=10^{-3}$. The classical quantity $R_C$ is also included for comparison.

    For the case $qE_0/m^2=1$, the approximate solution $\Delta E_Q$ consistently undervalues $\delta E_Q$ for both profiles. Given the form of the perturbative expansion $\Delta E_Q = \delta E_Q + \mathcal{O}(\delta E_Q^2)$, the extent to which $\Delta E_Q \neq \delta E_Q$ directly reflects the impact of higher-order perturbative terms. The early time regime for which growth in $R_Q$ was investigated in \cite{PlaNewsomeAnderson} can be adequately attenuated at $qt=5$, at which point the relative difference between $R_Q(qt=5)$ and $\Bar{R}_Q(qt=5)$ is of order $10^{-1}$ for both profiles. For $qE_0/m^2=10^3$, the same relative difference is of order $10^{-2}$ for the asymptotically constant profile and of order $10^{-3}$ for the Sauter pulse profile.

    Numerical results were also obtained for the cases $qE_0/m^2=0.5, 2, 10, 100$. The results provide evidence that if $qE_0/m^2 \gg 1$, then $\Delta E_Q \approx \delta E_Q$ for an extended period of time, whereas $\Delta E_Q \approx \delta E_Q$ only for relatively early times if $qE_0/m^2 \sim 1$. Since the validity analysis in~\cite{PlaNewsomeAnderson} was for early times, the conclusion in that paper that the semiclassical approximation breaks down for both the asymptotically constant and Sauter pulse profiles if $qE_0/m^2 \sim 1$ is verified here. The conclusion that the criterion used for the validity of the semiclassical approximation is not violated  at eary times for large values of $qE_0/m^2$ is also verified.

    What is also indicated by the above results is that over an extended period of time, it is likely that all solutions to the linear response equation $\eqref{fullLRE}$ are unstable for $qE_0/m^2 < \infty$, and only in the true massless limit $\eqref{LREmassless}$ is pure harmonic motion achieved and no instability present in $R_Q$. However, consideration of sufficiently late times is not physically realistic given the two major simplifications made of one spatial degree of freedom and ignoring spin $\sfrac{1}{2}$ particle interactions. Major modifications to the backreaction process and its linear response to perturbations are expected to occur if one, or both, of these limitations are relaxed.

\subsection{Impact of Quantum Fluctuations on Solutions to the Linear Response Equation}
\label{JJNum}

    The question of whether or not quantum fluctuations associated with the current density two-point function $\la [ J_Q(t,x),J_Q(t',x')]\ra$ are responsible for the instability observed in the solutions of the linear response equation \eqref{LRE}, and hence the validity of the semiclassical approximation, will now be addressed. For simplicity, the asymptotically constant classical source \eqref{Jasymp} will be considered here. However, similar qualitative results hold for the Sauter pulse classical source \eqref{JSauter}.

    From \eqref{masslesscomp}, one can see that $\delta \langle J_Q \rangle_\mathrm{ren}$ can be written in terms of its massless limit plus a non-local term. In the massless limit, one can see from \eqref{LREmassless} that $\la [ J_Q(t,x),J_Q(t',x')]\ra = 0$. This results in bounded harmonic behavior, leading to a constant value for $\bar{R}_Q$ in \eqref{RqdeltaE}. Thus for $qE_0/m^2 < \infty$, if there is growth in $\bar{R}_Q$ it must be driven by the non-local term, which depends in part on $\la [ J_Q(t,x),J_Q(t',x')]\ra$.

    The non-local term introduces a memory effect in the evolution of perturbations in the background field $\delta A$, the behavior of which depends not only on its current value at time $t$, but also on its entire history for past times $t'$. Initially, $\delta A(t'=0)=0$, but as $\delta A(t)$ evolves according to \eqref{LRE}, the time-dependent integration $\int_0^t dt'$ over the product of both $\int dx' \la [J_Q(t,x),J_Q(t',x')]\ra$ and $\delta A(t')$ accumulates contributions from the current density two-point function. Additionally, the quantity $\delta A(t')$ itself depends on the two-point function for all past times $t'$, creating a feedback loop where earlier perturbations in $\delta A(t')$ exert a delayed influence on $\delta A(t)$. As the upper limit of integration grows with $t$, the system continuously incorporates the effects of quantum fluctuations associated with the spin $\sfrac{1}{2}$ field, causing the current density two-point function to act as a type of amplifying source term. This drives the growth in $\delta A(t)$ leading to instability due to the influence of quantum fluctuations.

    The rate of this growth depends on the relative strength of $qE_0/m^2$. For $qE_0/m^2 \gg 1$, the time-dependent modes $h^{(I,II)}_k$ approach $\eqref{h1h2massless}$, thereby reducing the contribution of $\la [ J_Q(t,x),J_Q(t',x')]\ra$. This implies that quantum fluctuations as measured by this particular two-point correlation function diminish as the field strength grows, which is consistent with the behavior discusses previously for the massless case. However, one expects particle production to increase in this limit, emphasizing the point that the criterion for the validity of the semiclassical approximation used here is a necessary, but not sufficient condition. As the critical scale $qE_0/m^2 \sim 1$ is approached, the contribution of $\la [ J_Q(t,x),J_Q(t',x')]\ra$ is expected to be significant, indicating that the associated quantum fluctuations are large.

\begin{figure}[h]
    \centering
    \includegraphics[scale=0.4]{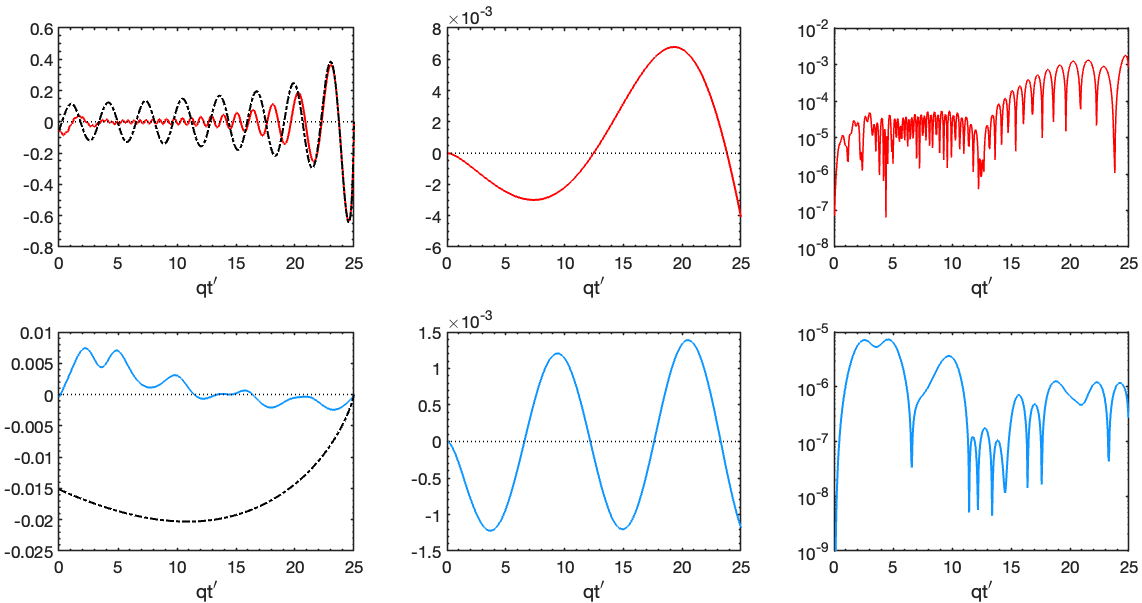} 
    \caption{The contents of the temporal integrand present in $\eqref{deltaJqrenorm}$ with perturbed classical source $\eqref{deltaJclass}$ are shown for cases $qE_0/m^2=1$ (top row) and $qE_0/m^2=10^3$ (bottom row). For both cases, $E_0/q=1$ so that $m^2/q^2=1$ for the top row plots and $m^2/q^2=10^{-3}$ for the bottom row plots. The spatial integral of the current density two-point function $-\frac{i}{q^3} \int_{-\infty}^\infty dx' \la [ J_Q(t,x),J_Q(t',x')]\ra$ (left), the perturbation in the gauge field $\delta A(t')$ (center), and the magnitude of their product $-\frac{i}{q^3} \int_{-\infty}^\infty dx' \la [ J_Q(t,x),J_Q(t',x')]\ra \delta A(t')$ (right) are plotted as a function of the past time $qt'$ up to a choice of current time $qt=25$. The current density two-point function in the absence of a background field (black dot-dashed curve) in \eqref{JJNoField} is included for comparison. Included for comparison, the dotted horizontal curve indicates when the corresponding quantity being plotted is zero.}
    \label{Fig5}
\end{figure}

    These features can be seen numerically in Fig.~\ref{Fig5}, where the contents of the temporal integrand present in $\eqref{deltaJqrenorm}$ are shown as a function of the past time $t'$ up to the current time $t$, which is chosen to be $qt=25$. For both characteristic cases $qE_0/m^2$ shown in the plots of the spatial integral of the current density two-point function, both situations in which the background field $E_0\neq 0$ as well as $E_0=0$ prior to activation are included.
    
    For $qE_0/m^2=1$, the spatial integral of the current density two-point function has an amplitude of oscillation which grows rapidly to a maximum as $t'\to t$ before terminating to zero when $t'=t$. This property is shared by both cases where $E_0=1$ and $E_0=0$, with there being good agreement between the two. However, for earlier values of $t'$ the amplitude of oscillations for $E_0=1$ are damped significantly with an increase in the frequency relative to $E_0=0$ occurring, resulting in a smaller contribution from earlier times for the integral over time $t'$. The amplitude of the quantity $\delta A(t')$ also grows significantly over the same time scale, which is a consequence of past contributions from the current density two-point function driving this growth. The magnitude of their product therefore results in rapid growth as $t'\to t$ which, when integrated in time $t'$ up to the current time $t$, drives the instability seen for $\bar{R}_Q$ in Fig.~\ref{Fig3}.
    
    For $qE_0/m^2=10^3$, the spatial integral of the current density two-point function is characterized by sporadic oscillations, with a relative amplitude approximately two orders of magnitude smaller than the critical case. Also, the maximum amplitude occurs for much earlier times $t'$, with damping occurring as $t' \to t$. Furthermore, it is only when $|t'-t|\ll 1$ that there is any agreement between the $E_0=1$ and $E_0=0$ cases, with the latter case having oscillations which take place over a much longer timescale. Consequently, the amplitude of $\delta A(t')$ does not grow significantly over the timescale considered, maintaining an approximately harmonic behavior. The integration of their product therefore yields a minimal contribution to the linear response equation, which in turn leads to much slower growth in $\bar{R}_Q$.

\section{Discussion and Conclusions}

    A linear response analysis has been conducted to investigate the validity of the semiclassical approximation to quantum electrodynamics in 1+1 dimensions for models in which pair production occurs due to the presence of a sufficiently strong electric field. A quantized massive spin $\sfrac{1}{2}$ field was considered which couples to a spatially homogeneous background electric field generated by a classical and conserved external source. Two classical current profiles were used which generate electric fields that are initially zero, as would be expected in a laboratory setting. The first involved the sudden activation of the current, with a corresponding electric field which asymptotically approaches a constant value. The second was the Sauter pulse, generated by a current that forms a smooth, time-dependent electric field pulse which is significant only over a finite time interval.

    Numerical results for the solutions to the linear response equation have been presented for both the critical threshold for pair production $qE_0/m^2 = 1$ as well as for $qE_0/m^2 =10^3$ where significant pair production occurs. An analytic solution to the linear response equation was found in the massless limit where $q E_0 /m^2 \to \infty$ for the asymptotically constant classical profile.

    A method of approximating homogeneous solutions to the linear response equation for semiclassical electrodynamics was utilized in \cite{PlaNewsomeAnderson}. It involves computing the difference between two solutions to the semiclassical backreaction equation which have similar initial conditions and was conjectured to be valid at sufficiently early times. That conjecture has been tested here for both classical source terms. For the critical threshold for pair production $q E_0/m^2 = 1$, it was found that the approximate solutions are in good agreement with the exact numerical solutions only for very early times.  For the much larger value $q E_0/m^2 = 10^3$, the agreement is significantly better at early times and extends to much later times.  In the massless limit with  $q E_0/m^2 \to \infty$, there is exact agreement between the difference between two solutions and the exact solution to the linear response equation for all times.

    As a result, the conclusions regarding the validity of the semiclassical approximation in~\cite{PlaNewsomeAnderson} have been verified by the analysis here. In particular, the solutions to the linear response equation, as measured by $\bar{R}_Q$ in \eqref{RqdeltaE}, exhibit significant growth at relatively early times for the critical case $q E_0/m^2=1$, indicating the validity criterion is violated and the semiclassical approximation breaks down. For considerably larger values of $qE_0/m^2$, the solutions to the linear response equation do not grow significantly at early times and the criterion is satisfied. However, the solutions do exhibit significant relative growth at sufficiently late times, suggesting that for all scales $qE_0/m^2 < \infty$ there will always be an instability present. These later times are not physically realistic for the 1+1 dimensional model considered here which neglects interactions between the produced particles. In the massless limit where $qE_0/m^2 \to \infty$, the solutions to the linear response equation do not grow in time so the criterion is never violated. In the large mass limit where $qE_0/m^2 \to 0$ for fixed $E_0$, particle production does not occur and the behavior of the electric field can be predicted by classical electrodynamics. If one extends to 3+1 dimensions, and/or if particle interactions are considered, then it is expected that there will be significant modifications to the behavior of the linear response equation. This is a subject of future investigation.

    The relationship between the growth of the solutions to the linear response equation and quantum fluctuations as measured by the retarded two-point correlation function for the current density has also been investigated in detail. The criterion for the validity of the semiclassical approximation used here assumes that quantum fluctuations associated with the current density two-point function are the mechanism by which relative growth occurs for perturbations of the semiclassical solutions. The question of whether this assumption is correct was addressed by examining \eqref{masslesscomp}, which gives the perturbation of the current density in terms of the perturbation in the massless case plus a non-local integral which contains the two-point correlation function. It was shown that the latter drives the growth in solutions since, by itself, the perturbed current in the massless case never causes the amplitudes of the solutions to increase in time.

    This was investigated further by considering the behavior of the retarded two-point function for the current density as a function of time in the case where there is no electric field, in the critical case with $q E_0/m^2=1$ and in the case $q E_0/m^2=10^3$. The details differ significantly between the three cases, but one thing that is true for all three is that for a given value of $q E_0/m^2$, the maximum value of the two-point correlation function does not grow significantly in time, even when solutions to the linear response equation do. It was also found that the two-point function does not vanish in the limit that the electric field does, even though the quantum current $\la J_Q \ra$ does.

    If the electric field is zero, then for fixed values of the current time parameter $t$, the spatial integral over the two-point function oscillates with a constant frequency and increasing amplitude as a function of the past time parameter $t'$. For $qE_0/m^2 = 1$, the behavior of this quantity for $t'$ relatively close to $t$ is the same as when $E_0=0$. However, for earlier values of $t'$ the oscillations are damped significantly. As a result, the contribution to the non-local term in \eqref{deltaJqrenorm} which involves the product of $\delta A(t')$ and this quantity, is heavily weighted towards the current time $t$, although the integrand vanishes in the limit $t' \to t$. For $qE_0/m^2=10^3$, the largest amplitude oscillations in the spatial integral of the two-point function come from significantly earlier times. This has the effect of slowing down the growth of solutions to the linear response equation.

    While these results specifically pertain to the semiclassical approximation to quantum electrodynamics in 1+1 dimensions for a spatially homogeneous electric field, it is very likely that they will generalize to other cases. In particular, it is often much easier to compute the difference between two solutions to the semiclassical backreaction equation with similar initial conditions than it is to solve the actual linear response equation. This will make it much easier to study the validity of the semiclassical approximation in other applications such as early universe cosmology.

\section*{Acknowledgments}
    We would like to thank Silvia Pla and Jose Navarro-Salas for helpful discussions. We would also like to thank Silvia Pla for sharing her numerical code to solve the semiclassical backreaction equations. 
    I. M. N. would like to thank Kaitlin Hill for helpful discussions. This work was supported in part by the National Science Foundation under Grants No. PHY-1912584 and PHY-2309186 to Wake Forest University. Some of the numerical work was done using the WFU DEAC Cluster; we thank the WFU Provost's Office and Information Systems Department for their generous support.

\section*{Data Availability}

    The data that support the findings of this article, and the numerical code from which it was generated, are openly available \cite{repo1,repo2}, embargo periods may apply.

\appendix

\section{Linear Response Equation Numerical Method for Homogeneous Perturbations}

    The linear response equation $\eqref{LRE}$ is a second order integro-differential equation governing the time evolution of homogeneous perturbations to the gauge field $\delta A(t)$, and by extension the electric field $\delta E(t)$, in the presence of sources $\delta J_C$ and $\delta \la J_Q \ra_\mathrm{ren}$. For numerical purposes, $\eqref{LRE}$ can be separated into two first order equations as
\bes
    \begin{flalign}
        \frac{d}{dt}\delta A(t) &= \chi(t) \quad , \\
        \frac{d}{dt}\chi(t) &= \delta J_C(t) + \delta \la J_Q (t) \ra_\mathrm{ren} \quad ,
    \end{flalign} \label{LRE2eqns}
\ees
    where it is understood that $\delta E(t) = -\chi (t)$. A $4^\mathrm{th}$ order Runge Kutta method was implemented to solve the system of equations $\eqref{LRE2eqns}$, iterating through the $i^\mathrm{th}$ value of the solution associated with the current time $t_i$ using
\bes
    \begin{flalign}
        \delta A_{i+1} &= \delta A_i + \frac{h}{6}\bigg(k_1+2k_2+2k_3+k_4 \bigg) \quad , \\
        \chi_{i+1} &= \chi_i + \frac{h}{6}\bigg(\ell_1+2\ell_2+2\ell_3+\ell_4 \bigg) \quad .
    \end{flalign}
\ees
    Here $h=\Delta t$, for a chosen timestep $\Delta t$. The relevant contributions are
\bes
    \begin{flalign}
        k_1 &= \chi_i \quad , \\
        k_2 &= \chi_i + \frac{\ell_1}{2} \quad , \\
        k_3 &= \chi_i + \frac{\ell_2}{2} \quad , \\
        k_4 &= \chi_i + \ell_3 \quad ,
    \end{flalign} \label{kays}
\ees
and
\bes
    \begin{flalign}
        \ell_1 &= \delta J_C(t_i) + \delta \left\la J_Q\left(t_i,\delta A_i\right) \right\ra_\mathrm{ren}  \quad , \\
        \ell_2 &= \delta J_C\left(t_i + \frac{h}{2}\right) + \delta \left\la J_Q\left(t_i + \frac{h}{2}, \delta A_i + \frac{k_1}{2}\right) \right\ra_\mathrm{ren} \quad , \\
        \ell_3 &= \delta J_C\left(t_i + \frac{h}{2}\right) + \delta \left\la J_Q\left(t_i + \frac{h}{2}, \delta A_i + \frac{k_2}{2}\right) \right\ra_\mathrm{ren} \quad , \\
        \ell_4 &= \delta J_C\left(t_i + h\right) + \delta \left\la J_Q\left(t_i + h, \delta A_i + k_3\right) \right\ra_\mathrm{ren} \quad .
    \end{flalign} \label{ells}
\ees
    From $\eqref{deltaJqrenorm}$, the quantum source perturbation $\delta \la J_Q \ra_\mathrm{ren}$ present in $\eqref{ells}$ can be expanded as
\bes
    \begin{flalign}
        \delta \la J_Q(t_i,\delta A_i) \ra _\textnormal{ren} &= -\frac{q^2}{\pi} \delta A_i + i \int_{t_0}^{t_i} dt' \int_{-\infty}^{\infty} dx' \, \la [J_Q(t_i,x) , J_Q(t',x')] \ra \, \delta A(t') \quad , \\
        \delta \left\la J_Q\left(t_i + \frac{h}{2}, \delta A_i + \frac{k_1}{2}\right) \right\ra_\mathrm{ren} &= -\frac{q^2}{\pi} \left( \delta A_i + \frac{k_1}{2} \right) \nonumber \\
        & + i \int_{t_0}^{t_i + \frac{h}{2}} dt' \int_{-\infty}^{\infty} dx' \, \left\la \left[J_Q\left(t_i + \frac{h}{2},x\right) , J_Q(t',x')\right] \right\ra \, \delta A(t') \, , \\
        \delta \left\la J_Q\left(t_i + \frac{h}{2}, \delta A_i + \frac{k_2}{2}\right) \right\ra_\mathrm{ren} &= -\frac{q^2}{\pi} \left( \delta A_i + \frac{k_2}{2} \right) \nonumber \\
        & + i \int_{t_0}^{t_i + \frac{h}{2}} dt' \int_{-\infty}^{\infty} dx' \, \left\la \left[J_Q\left(t_i + \frac{h}{2},x\right) , J_Q(t',x')\right] \right\ra \, \delta A(t') \, , \\
        \delta \la J_Q(t_i+h,\delta A_i+k_3) \ra _\textnormal{ren} &= -\frac{q^2}{\pi} \left(\delta A_i + k_3\right) \nonumber \\
        &+ i \int_{t_0}^{t_i+h} dt' \int_{-\infty}^{\infty} dx' \, \la [J_Q(t_i+h,x) , J_Q(t',x')] \ra \, \delta A(t') \quad .
    \end{flalign} \label{deltaJqrenormNum}
\ees

    For spatially homogeneous perturbations, the retarded two-point function for the current density will have the general property
\be
    \int_{-\infty}^{\infty} dx' \, \la [J_Q(t,x) , J_Q(t',x')] \ra \sim \int_{-\infty}^{\infty} dk \, g_k(t,t') \quad . \label{JJgen}
\ee
    where the function $g_k(t,t')$ depends on the relevant mode functions, which in turn depend on the background field. The relevant terms in $\eqref{deltaJqrenormNum}$ can be found from $\eqref{twopoint1}$. For the numerical results presented in this paper, the $k$-integral in $\eqref{JJgen}$ was computed using Simpson's $\sfrac{1}{3}$ rule.

    The initial conditions $\delta A(t=t_0) = 0$ and $\delta \dot{A}(t=t_0) = 0$ are the only available pieces of information one has from the outset, and therefore a left-handed Riemann sum method can be used to approximate the integral over $t'$ present in $\eqref{deltaJqrenormNum}$. The relevant integrals are
\bes
    \begin{flalign}
        &\int_{t_0}^{t_i} dt' \int_{-\infty}^{\infty} dx' \, \la [J_Q(t_i,x) , J_Q(t',x')] \ra \, \delta A(t') \nonumber \\
        & \qquad \qquad \qquad \approx \sum_{j=0}^{\frac{(t_i - t_0)}{\Delta t'}-1} \Delta t' \int_{-\infty}^{\infty} dx' \, \la [J_Q(t_i,x) , J_Q(t_j',x')] \ra \, \delta A(t_j') \quad , \\
        &\int_{t_0}^{t_i + \frac{h}{2}} dt' \int_{-\infty}^{\infty} dx' \, \left\la \left[J_Q\left(t_i + \frac{h}{2},x\right) , J_Q(t',x')\right] \right\ra \, \delta A(t') \nonumber \\
        & \qquad \qquad \qquad \approx \sum_{j=0}^{\frac{(t_i + h/2 - t_0)}{\Delta t'}-1} \Delta t' \int_{-\infty}^{\infty} dx' \, \left\la \left[J_Q\left(t_i + \frac{h}{2},x\right) , J_Q(t_j',x')\right] \right\ra \, \delta A(t_j') \quad ,  \label{halfstep} \\
        &\int_{t_0}^{t_i+h} dt' \int_{-\infty}^{\infty} dx' \, \la [J_Q(t_i+h,x) , J_Q(t',x')] \ra \, \delta A(t') \nonumber \\
        & \qquad \qquad \qquad \approx \sum_{j=0}^{\frac{(t_i + h - t_0)}{\Delta t'}-1} \Delta t' \int_{-\infty}^{\infty} dx' \, \la [J_Q(t_i+h,x) , J_Q(t_j',x')] \ra \, \delta A(t_j') \quad . \label{fullstep}
    \end{flalign} \label{JJsteps}
\ees
    However, the $4^\mathrm{th}$ order Runge-Kutta algorithm requires data, in part, for the solution $\delta A$ at a future half time step $t_i+h/2$ and a full time step $t_i+h$ as indicated by the upper limit of integration seen in $\eqref{halfstep}$ and $\eqref{fullstep}$. In terms of the left-handed Riemann sum method of approximating the $t'$-integral, this is data required of $\delta A_j$ for the index values $j=\frac{(t_i + h/2 - t_0)}{\Delta t'}-1$ and $j=\frac{(t_i +h - t_0)}{\Delta t'}-1$, respectively. Therefore, a method to estimate what these values for the linear response equation solutions $\delta A$ would be for the relevant future timesteps involves forward interpolation, to linear order, as
\bes
    \begin{flalign}
        \delta A \left(t' = t_i + \frac{h}{2} \right) &\approx \delta A(t_i) + \frac{h}{2} \chi(t_i) \\
        \delta A \left(t' = t_i + h \right) &\approx \delta A(t_i) + h \, \chi(t_i)
    \end{flalign}
\ees
    This completes the necessary elements required to obtain solutions to the linear response equation $\eqref{LRE}$ for homogeneous perturbations.

    The above method has been shown to yield accurate results for a test integro-differential equation of the form
\be
    \frac{d^2}{dt^2}f(t) = J - f(t) - \int_{t_0}^t dt' (t-t') f(t') \quad . \label{test}
\ee
    Here $J$ is taken to be a constant. With $t_0=0$, such that the initial conditions are $f(t=0)=0$ and $\dot{f}(t=0)=0$, the analytic solution to $\eqref{test}$ is
\be
    f(t) = \frac{2}{\sqrt{3}}J \sin{\left(\frac{\sqrt{3}}{2}t \right)} \sinh{\left(\frac{t}{2}\right)} \quad .
\ee
    With a step size $\Delta t' = 10^{-3}$, the relative difference between the exact and numerical solution was found to be of order $10^{-15}$, providing evidence of this method's accuracy.

\section{Comparison of Two Polarization Tensors}

    For the semiclassical investigation contained in this paper, a retarded polarization tensor can be defined as
\be
    \Pi^\mathrm{ret}_{\mu \nu}(x,x') = i \theta(t-t') \la [J_\mu(x),J_\nu(x')]\ra \quad , \label{polar1}
\ee
    such that the perturbed quantum source term \eqref{deltaJqrenorm} present in the linear response equation \eqref{LRE} can be written as
\be
    \delta \la J_Q(t) \ra _\textnormal{ren} = -\frac{q^2}{\pi} \delta A(t) + \int_{-\infty}^\infty dt' \int_{-\infty}^{\infty} dx' \, \Pi^\mathrm{ret}_{11}(x,x') \, \delta A(t') \quad .
\ee
    A different approach to studying the Schwinger effect involves using scattering theory in perturbative quantum electrodynamics \cite{Gitman,Fradkin,Shvartsman,Soviet}. In this case the polarization tensor is defined as \cite{Soviet}
\be
    \bar{\Pi}_{\mu \nu}(x,x') = i q^2 \, \mathrm{Tr}\left\{ \gamma_\mu S(x,x') \gamma_\nu S(x',x) \right\} \quad , \label{polar2}
\ee
    where $S(x,x')$ is the Feynman propagator.
    
    However, \eqref{polar2} can be related to a two-point function for the current density similar in character to \eqref{polar1}. To see this, first consider the following
\be 
    \Pi_{\mu \nu}(x,x') = i \la T\left( J_\mu(x) J_\nu(x') \right) \ra \quad . \label{polar3}
\ee
    Substituting \eqref{J2} into the time-ordered product in \eqref{polar3} yields the following
\be
    T\left( J_\mu(x) J_\nu(x') \right) = q^2 \sum_{a,b,c,d} \left(\gamma_\mu\right)_{ab} \, \left(\gamma_\nu\right)_{cd} \, T\left( \bar{\psi}_a(x) \, \psi_b(x) \, \bar{\psi}_c(x') \, \psi_d(x') \right) \quad .
\ee
    Here, the subscripts $\{a,b,c,d\}$ correspond to spinor indices. Utilizing Wick's theorem for time-ordered products of fermion fields, one has
\begin{flalign}
    T\left( \bar{\psi}_a(x) \, \psi_b(x) \, \bar{\psi}_c(x') \, \psi_d(x') \right) &= \, :\bar{\psi}_a(x) \, \psi_b(x) \, \bar{\psi}_c(x') \, \psi_d(x'): \nonumber \\
    & \, + : \bar{\psi}_a \overbracket{(x) \, \psi_b}(x) \, \bar{\psi}_c(x') \, \psi_d(x'): \nonumber \\
    & \, + :\bar{\psi}_a \overbracket{(x) \, \psi_b(x) \, \bar{\psi}_c(x') \, \psi_d}(x'): \nonumber \\
    & \, + :\bar{\psi}_a(x) \, \psi_b \overbracket{(x) \, \bar{\psi}_c}(x') \, \psi_d(x'): \nonumber \\
    & \, + :\bar{\psi}_a(x) \, \psi_b(x) \, \bar{\psi}_c \overbracket{(x') \, \psi_d}(x'): \nonumber \\
    & \, + :\bar{\psi}_a \overbracket{(x) \, \psi_b}(x) \, \bar{\psi}_c \overbracket{(x') \, \psi_d}(x'): \nonumber \\
    & \, + :\bar{\psi}_a \overbracket{(x) \, \psi_b \overbracket{(x) \, \bar{\psi}_c}(x') \, \psi_d}(x'): \quad , \label{timeorder1}
\end{flalign}
    where a term such as $\bar{\psi}_a \overbracket{(x) \, \psi_b}(x)$ represents a contraction between the fields $\bar{\psi}_a(x)$ and $\psi_b(x)$ and is given by
\be
    \bar{\psi}_a \overbracket{(x) \, \psi_b}(x) = \bra{0} T\left( \bar{\psi}_a(x) \, \psi_b(x)\right) \ket{0} \quad .
\ee
    When considering the vacuum state expectation value of \eqref{timeorder1}, one is left with
\begin{flalign}
    \left\langle T\left( \bar{\psi}_a(x) \, \psi_b(x) \, \bar{\psi}_c(x') \, \psi_d(x') \right) \right\rangle &= \left\langle T\left(\bar{\psi}_a(x) \, \psi_b(x) \right)\right\rangle \, \left\langle T\left(\bar{\psi}_c(x') \, \psi_d(x') \right)\right\rangle \nonumber \\
    & + \left\langle T\left(\bar{\psi}_a(x) \, \psi_d(x') \right)\right\rangle \, \left\langle T\left(\psi_b(x) \, \bar{\psi}_c(x') \right)\right\rangle \quad . \label{timeorder2}
\end{flalign}
    These remaining terms are related to the Feynman propagators defined in \cite{Soviet}, which are given by
\be
    S(x,x') = i \la T\left( \psi(x) \bar{\psi}(x') \right) \ra \quad .
\ee
    In order to make contact with this form of the propagator, one can utilize the equal time anticommutation relations
\be
    \{ \psi_a(t, \Vec{x}) , \bar{\psi}_b(t,\Vec{x}') \} = ( \gamma^0 )_{ab} \, \delta^{(3)}(\Vec{x}-\Vec{x}') \quad ,
\ee
    to re-arrange the first three time-ordered products on the right-hand-side of \eqref{timeorder2}. Recall the time-ordering operator is given by
\be
    T\left( \bar{\psi}_a(x) \psi_b(x') \right) = \bigg\{
    \begin{array}{c}
         \bar{\psi}_a(x) \psi_b(x') \quad \, \, \, \, \, \mathrm{if} \quad t>t' \\
        -\psi_b(x')\bar{\psi}_a(x) \quad \mathrm{if} \quad t<t'
    \end{array} \quad .
\ee
    For the case when the spacetime points are equal inside the time-ordering operation, which occurs for the first two time-ordered products in \eqref{timeorder2}, one is left with the relation
\be
    \la \bar{\psi}_a(x) \psi_b(x) \ra = - \la \psi_b(x) \bar{\psi}_a(x) \ra + (\gamma^0)_{ab} \, \delta^{(3)}(0) \quad .
\ee
    For the case when the spacetime points are not equal, and assuming $t<t'$, one simply has
\be 
    \la T\left( \bar{\psi}_a(x) \psi_b(x') \right) \ra = - \la T\left( \psi_b(x') \bar{\psi}_a(x) \right) \ra \quad .
\ee
    Therefore, \eqref{polar3} can be expressed as
\begin{flalign}
    \Pi_{\mu \nu}(x,x') 
    &= i \, q^2 \bigg[i \sum_{a,b} \left(\gamma_\mu\right)_{ab} S_{ba}(x,x) + \sum_{a,b} \left(\gamma_\mu\right)_{ab}(\gamma^0)_{ab} \, \delta^{(3)}(0) \bigg] \nonumber \\
     &\quad \, \, \, \times \bigg[ i \sum_{c,d} \left(\gamma_\nu\right)_{cd} S_{dc}(x',x') + \sum_{c,d} \left(\gamma_\nu\right)_{cd} (\gamma^0)_{cd} \, \delta^{(3)}(0) \bigg] \nonumber \\
     &\quad \, \, \, + i \, q^2 \sum_{a,b,c,d} \left(\gamma_\mu\right)_{ab} S_{bc}(x,x') \, \left(\gamma_\nu\right)_{cd} S_{da}(x',x) \nonumber \\
    &= i \, q^2 \bigg[ i \, \mathrm{Tr}\left\{ \gamma_\mu S(x,x) \right\} + \mathrm{Tr}\left\{ \gamma_\mu \gamma^0 \right\} \delta^{(3)}(0) \bigg] \nonumber \\
    & \quad \, \, \, \times \bigg[ i \, \mathrm{Tr}\left\{ \gamma_\mu S(x',x') \right\} + \mathrm{Tr}\left\{ \gamma_\mu \gamma^0 \right\} \delta^{(3)}(0) \bigg] \nonumber \\
    & \quad \, \, \, + i \, q^2 \, \mathrm{Tr} \left\{ \gamma_\mu S(x,x') \, \gamma_\nu S(x',x) \right\} \quad , \label{polartrace}
\end{flalign}
    where the following trace relations were implemented
\bes
    \begin{flalign}
        \mathrm{Tr}\{ A B \} &= \sum_{a,b} A_{ab} B_{ba} \quad , \\
        \mathrm{Tr}\{ A B C D \} &= \sum_{a,b,c,d} A_{ab} B_{bc} C_{cd} D_{da} \quad .
    \end{flalign}
\ees
    The polarization tensor in \eqref{polartrace} is comprised of a sum of disconnected and connected quantities. In physical applications, one is typically concerned with the connected correlator because it encodes the interaction structure between fields. The disconnected part, while nonzero, often represents background contributions that are subtracted or ignored in defining physical observables. However, the vacuum state expectation value of the current density can be put into the form
\begin{flalign}
        \la J_\mu(x) \ra &= q \, \la \bar{\psi}(x) \gamma_\mu \psi(x) \ra \nonumber \\
        &= q \sum_{a,b} \left(\gamma_\mu\right)_{ab} \left\langle \bar{\psi}_a(x) \psi_b(x) \right\rangle \nonumber \\
        &= q \sum_{a,b} \left(\gamma_\mu\right)_{ab} \bigg[ -\left\langle \psi_b(x) \bar{\psi}_a(x) \right\rangle + (\gamma^0)_{ab} \, \delta^{(3)}(0) \bigg] \nonumber \\
        &= i \, q \sum_{a,b} \left(\gamma_\mu\right)_{ab} S_{ba}(x,x) + q \sum_{a,b} \left(\gamma_\mu\right)_{ab}(\gamma^0)_{ba} \, \delta^{(3)}(0) \nonumber \\
        &= i \, q \, \mathrm{Tr}\left\{ \gamma_\mu S(x,x) \right\} + q \, \mathrm{Tr}\left\{ \gamma_\mu \gamma^0 \right\} \delta^{(3)}(0) \quad ,
    \end{flalign}
    where in the fourth equality the property $(\gamma^0)_{ab}=(\gamma^0)_{ba}$ was used. Therefore, one can define the connected part of \eqref{polar3} as
\begin{flalign}
    \Pi_{\mu \nu}(x,x')_\mathrm{conn} &\equiv \Pi_{\mu \nu}(x,x') - i \la J_\mu(x) \ra \la J_\nu(x') \ra \quad , \label{polarAgain}
\end{flalign}
    which is equivalent to $\bar{\Pi}_{\mu \nu}(x,x')$ in \eqref{polar2}.

    Although the polarization tensors \eqref{polar1} and \eqref{polar3} are superficially similar, the physical context in which they are utilized is quite different. In the scattering theory framework, the electromagnetic field is quantized and treated as a nontrivial background, enabling the perturbative study of particle production processes and quantum corrections, such as vacuum polarization effects that modify the photon propagator. However, strong backreaction effects on the background field are not included in this formalism. In contrast, strong backreaction effects can be taken into account by solving the semiclassical backreaction equations. The linear response analysis used in this paper involves perturbations of solutions to the semiclassical backreaction equations.

\end{document}